\documentclass[%
reprint,
superscriptaddress,
nofootinbib,
 amsmath,amssymb,
prl,
floatfix,
]{revtex4-1}
\usepackage[dvipsnames]{xcolor}

\usepackage[utf8]{inputenc}
\usepackage{graphicx}
\usepackage{xcolor}
\usepackage{comment}
\usepackage{soul}
\usepackage{siunitx}
\usepackage{textgreek}
\usepackage[mathlines]{lineno}
\usepackage{tabularx}
\usepackage[utf8]{inputenc}

\usepackage{lipsum}  
\usepackage{enumitem}
\usepackage{xspace} 

\newcommand{\eV}{\ensuremath{\,\mathrm{eV}}\xspace}
\newcommand{\keV}{\ensuremath{\,\mathrm{keV}}\xspace}

\newcommand{\kevnr}{\ensuremath{\,{\mathrm{keV_{nr}}}}\xspace}
\newcommand{\kevt}{\ensuremath{\,{\mathrm{keV_{t}}}}\xspace}
\newcommand{\kevee}{\ensuremath{\,{\mathrm{keV_{ee}}}}\xspace}

\newcommand{\neh}{n_\mathrm{eh}}

\begin{document}

\title{First measurement of the nuclear--recoil ionization yield in silicon at 100\,eV
}


\author{M.F.~Albakry} \affiliation{Department of Physics \& Astronomy, University of British Columbia, Vancouver, BC V6T 1Z1, Canada}\affiliation{TRIUMF, Vancouver, BC V6T 2A3, Canada}
\author{I.~Alkhatib} \affiliation{Department of Physics, University of Toronto, Toronto, ON M5S 1A7, Canada}
\author{D.~Alonso} \affiliation{Instituto de F\'{\i}sica Te\'orica UAM/CSIC, Universidad Aut\'onoma de Madrid, 28049 Madrid, Spain}\affiliation{Instituto de F\'{\i}sica Te\'orica UAM-CSIC, Campus de Cantoblanco, 28049 Madrid, Spain}
\author{D.W.P.~Amaral} \affiliation{Department of Physics, Durham University, Durham DH1 3LE, UK}
\author{P.~An} \affiliation{Department of Physics, Duke University, Durham, NC 27708, USA}
\author{T.~Aralis} \affiliation{Division of Physics, Mathematics, \& Astronomy, California Institute of Technology, Pasadena, CA 91125, USA}
\author{T.~Aramaki} \affiliation{Department of Physics, Northeastern University, 360 Huntington Avenue, Boston, MA 02115, USA}
\author{I.J.~Arnquist} \affiliation{Pacific Northwest National Laboratory, Richland, WA 99352, USA}
\author{I.~Ataee~Langroudy} \affiliation{Department of Physics and Astronomy, and the Mitchell Institute for Fundamental Physics and Astronomy, Texas A\&M University, College Station, TX 77843, USA}
\author{E.~Azadbakht} \affiliation{Department of Physics and Astronomy, and the Mitchell Institute for Fundamental Physics and Astronomy, Texas A\&M University, College Station, TX 77843, USA}
% ...



\date{\today}

\begin{abstract}
We measured the nuclear--recoil ionization yield in silicon with a cryogenic phonon-sensitive gram-scale detector. Neutrons from a mono-energetic beam scatter off of the silicon nuclei at angles corresponding to energy depositions from 4\,keV down to 100\,eV, the lowest energy probed so far. The results show no sign of an ionization production threshold above 100\,eV. These results call for further investigation of the ionization yield theory and a comprehensive determination of the detector response function at energies below the keV scale. 

\end{abstract}
\maketitle


The identity of Dark Matter and determination of neutrino properties are problems at the forefront of physics beyond the Standard Model. Rare event searches focused on Dark Matter detection~\cite{abdelhameed2019first,edelweiss2018,agnese2017projected,cdex2020,sensei2019,damic2020} or Coherent Elastic Neutrino-Nucleus Scattering (CE$\nu$NS)~\cite{akimov2017observation,cabrera2012development,wong2005low,belov2015nugen,buck2017conus,billard2017coherent,agnolet2017background,aguilar2016connie} often detect products generated by a particle interacting in a target material, requiring a strong understanding of that material's response to energy depositions. Silicon is a commonly used target material. Particle interactions with the silicon nuclei or electrons generate free charge carriers, with nuclear recoils generating fewer charge carriers than electron recoils of the same energy. The ratio of charge carriers produced by nuclear and electron recoils, called the ionization yield $Y$, is crucial to understanding the response of such detectors, and is believed to be an intrinsic material property. Experimental measurements of $Y$ in silicon~\cite{sattler1965ionization,gerbier1990measurement,dougherty1992measurements} for nuclear recoils above 4\keV have been consistent with a model developed by Lindhard \textit{et al}.~\cite{lindhard1963integral}. At lower energies, typical in low-mass Dark Matter or CE$\nu$NS searches, measurements of $Y$ indicate a significant deviation from the Lindhard model~\cite{chavarria2016measurement,izraelevitch2017measurement, Villano:2022}. Recent modeling~\cite{Sorensen:2015df,Sarkis:2020ds} has focused on understanding the origin of these deviations. Furthermore, measurements of the low-energy yield in another commonly used semiconductor, germanium, have been inconsistent with each other~\cite{Collar2021yield,Albakry2022photoneutron,bonhomme2022geyield}. These observations create the need for a range of ionization yield measurements. Here, we present the result of an ionization yield measurement in silicon using data taken with a cryogenic detector~\cite{Romani:2018,ren:2020} as part of a neutron scattering experiment called IMPACT (Ionization Measurement with Phonons At Cryogenic Temperatures).

We perform the measurements using the Tandem accelerator at Triangle Universities Nuclear Laboratory. The accelerator produces a pulsed proton beam, which is directed onto a 100-nm-thick lithium fluoride (LiF) target. The target reaction $\mathrm{^7}$Li(p,n)$\mathrm{^7}$Be produces a neutron beam with a controllable bimodal energy after collimation~\cite{lee1991neutrons,Hanson:1949}. The ejected neutrons elastically scatter in the silicon detector, and neutrons from the higher energy mode are subsequently detected in liquid scintillator cells located at known scattering angles corresponding to six recoil energies between 0.1 and 3.9\kevnr.

By tuning the proton energy to slightly over the forward production threshold (1.881\,MeV) and selecting the forward-going neutrons, we produce a neutron beam at 55.7\keV with $\sim$1\keV spread. This allows us to exploit the 55.7\keV resonance in the neutron--silicon elastic-scattering cross section, making the measurement robust against small drifts in the neutron energy. At the beginning of the experiment, the neutron energy is measured to match the expected $
\sim$56\keV via the time-of-flight between the beam pickup monitor (BPM) immediately upstream from the target and a neutron detector at $0^{\circ}$.

The silicon detector is a 1-cm$^2$-square and 4-mm-thick SuperCDMS HVeV detector~\cite{Romani:2018,ren:2020} measuring the total phonon energy (denoted as $E_t$ with unit \kevt) generated following particle scattering. The detector is operated at $52$\,mK in an Adiabatic Demagnetization Refrigerator (ADR). A voltage bias can be applied across the 4-mm thickness, producing phonons from the accelerated charge carriers through the Neganov-Trofimov-Luke effect~\cite{neganov1985colorimetric, Luke:1988}. The total phonon energy $E_t$ is
\begin{equation}
    E_{t}=E_{r} + \neh \cdot eV,~\text{where}~\left\langle\neh\right\rangle = Y\cdot \frac{E_{r}}{\epsilon}.
    \label{eqn:totalphononenergy}
\end{equation}
The term $E_{r}$ is the recoil energy, $e$ the elementary electric charge, $V$ the substrate voltage bias, $\neh$ the number of electron-hole ($eh$) pairs  generated from this recoil, $\left\langle\neh\right\rangle$ its averaged number for an energy deposit of $E_r$, and $\epsilon=3.8$\eV~\cite{pehl1968accurate} the average energy per $eh$ in silicon for deposits $\gg$20~eV~\cite{Ramanathan:2020}. By operating without voltage bias (0\,V mode) we measure the recoil energy directly, whereas by applying a voltage bias (HV mode) our detector becomes sensitive to the ionization signal. This work presents an ionization yield measurement based on 11 days of data taken with a bias of 100\,V, with cross-checks performed in 0\,V mode.

We record the silicon detector data continuously at a sampling frequency of 1.5\,MS/s and use an offline trigger to identify energy depositions. The detector achieved a baseline energy resolution of $\sim$4.5\,$\mathrm{eV_{t}}$. A trigger threshold of 50\,$\mathrm{eV_{t}}$ is used to avoid near-threshold effects.
Calibration of the detector up to phonon energies of 120\kevt is accomplished with a laser and an $^{55}$Fe source~\cite{ren:2020}. Laser calibration data are taken daily to allow us to correct for gain variations caused by ADR thermal cycles.

For the secondary neutron detection, we use a total of 29 liquid scintillator cells filled with Eljen EJ-301 or EJ-309 coupled to Hamamatsu R7724 photo-electron multiplier tubes (PMT) arranged at angles corresponding to six different nuclear recoil energies. Twenty-six of the neutron detectors are mounted on two concentric rings of radii 29.4\,cm and 45.2\,cm, referenced as ring detectors. The rings are centered on the beam axis and are first placed at a distance of 86\,cm downstream from the silicon detector for 7 days to perform the measurement at 0.46 and 0.22\kevnr, and then at 131\,cm downstream for 4 days for 0.22 and 0.1\kevnr. The remaining three neutron detectors, referenced as the lone-wolf (LW) detectors, are each positioned to measure recoil energies of 0.75, 2.2 and 3.9\kevnr $\sim$20\,cm away from the silicon detector.
The scattered neutrons deposit a maximum of $\sim$50~keV in the liquid scintillator, equivalent to a 5~keV electron recoil (5\kevee) assuming a quenching factor of 10\%~\cite{awe2018liquid}.
The neutron detectors are calibrated daily against the Compton edge of $^{137}$Cs gamma rays.

The PMT and BPM waveforms are digitized simultaneously at a sampling frequency of 250\,MS/s following a hardware trigger window of $\sim$3\kevee to $\sim$40\kevee on each PMT channel. We expect negligible detection of the low-energy mode of neutrons from the beam~\cite{awe2018liquid}. To identify coincidence events between the silicon and the secondary neutron detectors, we synchronize the clock of the two data acquisition systems every minute.

To understand the expected recoil energy distributions for each secondary neutron detector, we simulate the experiment using \textsc{Geant4}~\cite{brun1993geant}-10.05.p01 with the \textit{Shielding} physics list. The simulation model includes the neutron beam collimator, the ADR, the silicon detector, and the scintillator cells of the secondary neutron detectors. Neutrons are emitted from a spot of 1-mm radius in a cone of $4^{\circ}$ half-angle from the location of the target. Angular dependence of the neutron kinematics from the emission is negligible for a collimator with a $3^{\circ}$ opening angle~\cite{lee1991neutrons}.

The silicon detector position is determined to within $\sim$1\,cm by scanning with a tightly collimated $^{57}$Co x-ray source on a translation stage. The detector position uncertainty translates into an uncertainty on the measured recoil energy. This uncertainty is negligible for the ring detectors but not for the LW detectors. To account for this, we linearly scale the simulated LW recoil energy spectra to the LW spectra measured in 0\,V mode. The scale factors, determined by a binned-likelihood minimization, are reported in Table~\ref{table:recoil_energy}. We adopt the uncertainties on these scale factors as a systematic uncertainty on the recoil energy measured by the LWs.

\begin{table}[htbp]
\begin{tabular}{ccc}
\hline\hline
Nominal $E_r$ & Measured $E_r$ & Scale factor \\ \hline
0.75 keV          & 0.89 keV               & 1.18$^{+0.03}_{-0.15}$         \\
2.00 keV          & 2.33 keV               & 1.16$^{+0.05}_{-0.14}$     \\
3.87 keV          & 3.91 keV               & 1.01$^{+0.01}_{-0.11}$         \\ \hline\hline
\end{tabular}
\caption{Recoil energy for the LW detectors. }
\label{table:recoil_energy}
\end{table}

The initial neutron energy distribution in the simulation is generated from a semi-analytical model of the proton beam interactions in the LiF target. Initial proton energies are sampled from a Gaussian distribution with a standard deviation of 2\keV~\cite{tunl-cdms:2019} and a mean value depending on the targeted neutron energy. The energy loss in LiF, as determined by TRIM~\cite{ziegler:2010}, can then be converted into a neutron energy following the kinematics outlined in Ref.~\cite{Hanson:1949}. We sample neutron energy distributions for targeted nominal beam energies ranging from 46 to 60\keV in 1\keV steps.

We use these simulations to evaluate the stability of the neutron beam energy hourly based on a fit of the spectrum of neutron energy depositions in the silicon detector between 4.0 and 8.2~keV. The lower bound is outside our region of interest for the ionization yield analysis, while the upper bound marks an energy region where neutron interactions dominate. Since the discrepancy between the Lindhard model and existing measurements is only 20\% in this energy range, we use it to convert the simulated recoil energies to total phonon energies. The neutron beam simulation matching the data best is chosen to represent the beam behavior for that hour. This method yields a spread in the best-fit beam energy of about 3\,keV, which we take to be the systematic uncertainty on the beam energy. 


We remove silicon detector events occurring on the tails of large energy depositions to ensure accuracy in energy reconstructions. Multiple energy depositions occurring close in time, referred to as pile-up events, pose challenges in energy estimations. We allow up to 2 pile-ups per event after demonstrating negligible biases in energy estimations using the ``matched-filter-integral" algorithm with built-in pileup corrections~\cite{ren:2020}. 
For neutron detector events, we remove pile-up events and select events below 10\kevee to reject photon events. Events where most energy is deposited within a single time bin are also removed, as these are inconsistent with the expected neutron pulse shape~\cite{kuchnir1968time}.

\begin{figure}[htbp]
    \centering
    \includegraphics[width=0.47\textwidth]{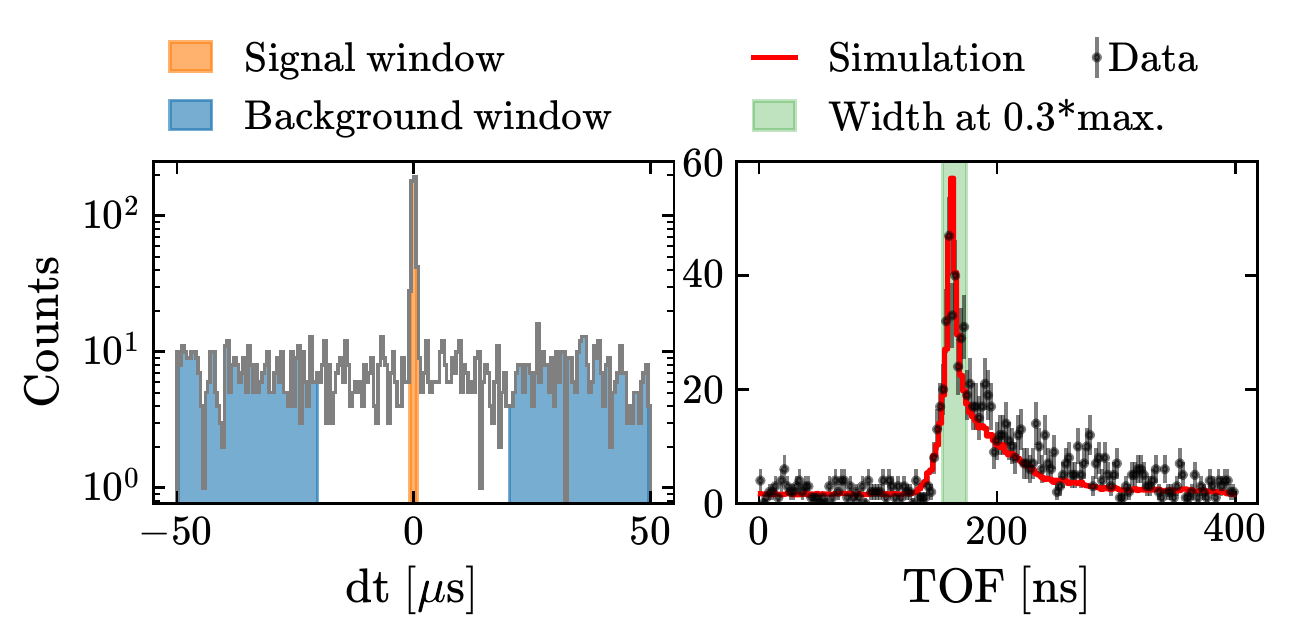}    
    \caption{Time difference between the silicon detector and neutron detectors ($dt$) and time difference between BPM and neutron detectors (TOF) for 890~eV$_{\mathrm{nr}}$ scattering. Left: $dt$ distribution. The signal window is marked in orange; the region corresponding to random coincidences used for background estimation is marked in blue. Right: TOF distribution. The selection window is marked in green.}
    \label{fig:dt_tof}
\end{figure}

We define a variable $dt$ as the time difference between any silicon detector event and neutron detector events. Pairs of events in the silicon and the neutron detectors with $|dt-dt_0|<0.9\,\mathrm{\mu s}$ are referred to as coincident events, in which $dt_0$ is the neutron time of flight between the two detectors. The $0.9\,\mathrm{\mu s}$ coincidence window corresponds to 3 standard deviations of the $dt$ timing resolution (see Fig.~\ref{fig:dt_tof}).

\begin{figure}[htbp]
    \centering
    \includegraphics[width=0.47\textwidth]{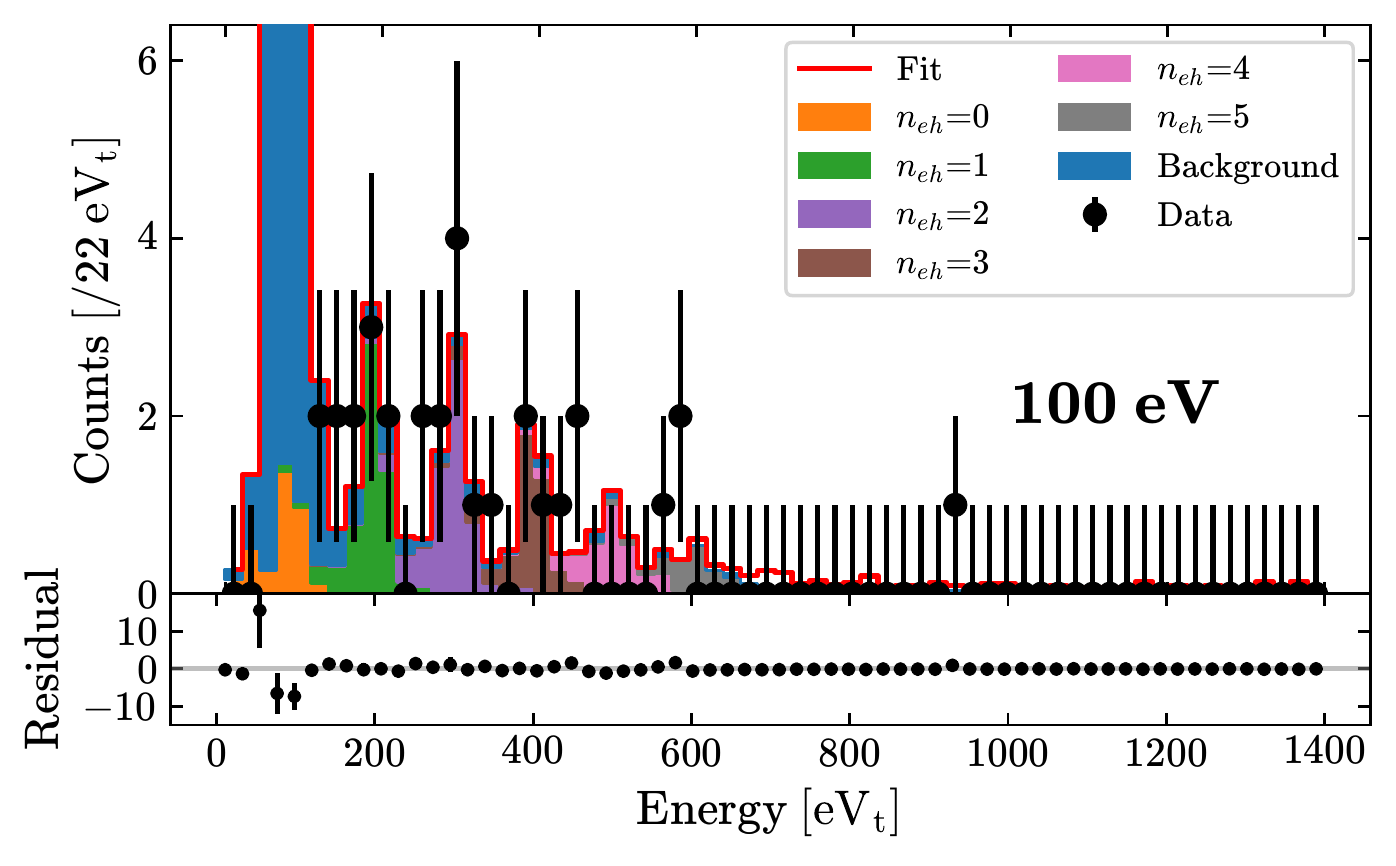}  
    \includegraphics[width=0.47\textwidth]{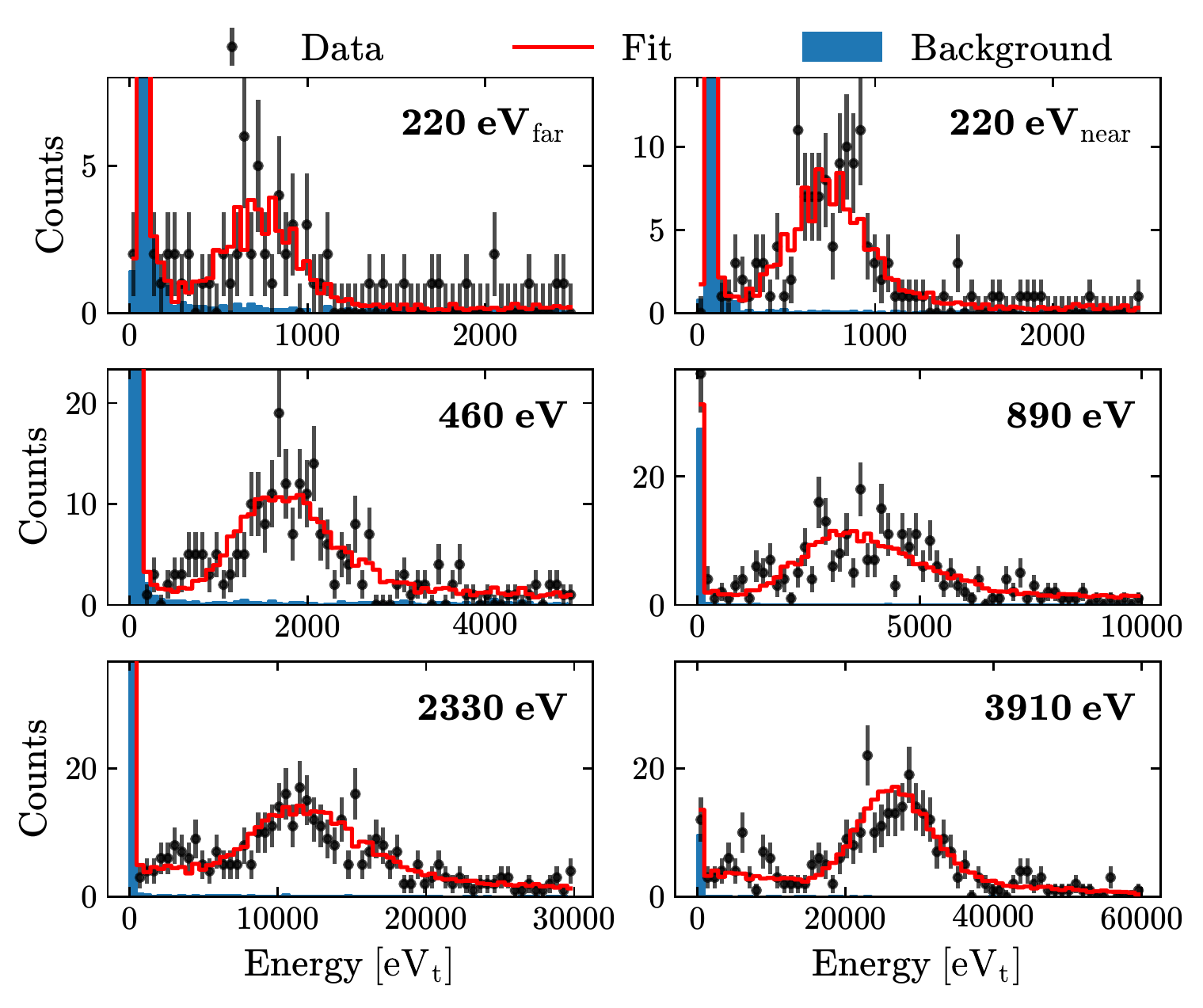}    
    \caption{Comparison of the fit model using the best-fit parameters to the data for all recoil energies. For the 100\,eV recoil energy (top panel), the integer number of $eh$ contributions to the model are shown colored for pair number up to $\neh=5$. Individual $eh$ pairs are not shown for other recoil energies because they are indistinguishable based on our energy resolution and the number of events. Blue regions are the estimated background.}
    \label{fig:spectra_all}
\end{figure}

To suppress background from random coincidences, we require the time difference between BPM and neutron detector (denoted as TOF) to be consistent with 55.7\keV neutrons. This corresponds to a peak in the TOF distribution (see Fig.~\ref{fig:dt_tof}). The simulated TOF is fitted to data with a time offset to account for a small discrepancy in the geometry. We select events with a window corresponding to a $\pm2\,$keV spread in the neutron energy. The energy spectra of all recoil energies after event selection are shown in Fig.~\ref{fig:spectra_all}.

Background events with no correlation between the neutron detector and the silicon detector form a flat distribution in $dt$. We estimate them with side-bands from 20 to 50\,$\mu$s before or after the coincidences, as shown in Fig.~\ref{fig:dt_tof}. The estimated background spectra are shown as the blue colored component in Fig.~\ref{fig:spectra_all}. 

We build a detector response model that transforms the simulated nuclear recoil energy spectra into the total phonon energy spectra. We parameterize $Y(E_r)$, the energy-dependent ionization yield, as a piecewise linear function characterized by the values at the recoil energies we measure. The function is fixed at low and high energies to be $Y(E_r=0\keV)=0$, and $Y(E_r=10 \keV)=0.3$ from the Lindhard model. For each simulated event with $E_r$, $\left\langle\neh\right\rangle$ is calculated with Eq.~\ref{eqn:totalphononenergy}. Then $\neh$ is sampled from a distribution with this mean and a variance ($\sigma^2(\neh)$) characterized by a Fano factor $F=\sigma^2(\neh)/\left\langle\neh\right\rangle$. Poission, binomial and negative binomial distributions are used to model $F$=1, $F$\textless1, and $F$\textgreater1, respectively. The nuclear-recoil Fano factor is unknown at this energy but at larger energies there is evidence that it can be $\gg 1$ \cite{Matheny2022fano}.
$\neh$ is further smeared to account for charge trapping (12.7\% probability) and impact ionization (0.6\% probability) in the detector~\cite{ren:2020}. An energy-dependent Gaussian distributed detector resolution (see Ref.~\cite{ren:2020}) is applied to the total phonon energy $E_t$ after conversion from $\neh$ with Eq.~\ref{eqn:totalphononenergy}.

We compare the simulated $E_t$ spectra to data after applying their normalization factors, denoted as $n$. Each simulated $E_t$ spectrum has three free parameters: $Y$, $F$, and $n$. For the 220\eV recoil energy point that is measured twice, we constrain the two fits to have the same yield and the same Fano factor. We sample these parameters simultaneously with a Markov-Chain Monte Carlo (MCMC) method using a binned-likelihood loss function implemented in the Bayesian Analysis Toolkit~\cite{caldwell2009bat}.

To improve the speed of convergence, we perform the fit in an iterative way. The first fit is for the ring detector parameters while keeping the yield at the LW detector energies fixed to the Izraelevitch result~\cite{izraelevitch2017measurement}. We then fit for the LW detector parameters while keeping the ring parameters at the previous result. Finally, the fit is rerun on the ring parameters with the LW parameters at the previous result. The results with the best-fit parameters is shown in Fig.~\ref{fig:spectra_all}.

\begin{table*}[htbp]
\def\arraystretch{1.3}
\begin{tabular}{ m{4em} m{6em} m{4em} m{10em} m{4em} m{4em} m{4em} m{4em} m{4em} m{6em}} 
\hline\hline
$E_{r}$ {[}$\kevnr${]} & Ionization yield $Y$ & Fano factor & Normalization & Stat. & Recoil energy & Beam energy & CT/II & TOF & Fano factor mismodeling \\\hline
0.10     & $0.102^{+0.034}_{-0.030}$ & $0.9_{-0.4}^{+0.7}$ & $28_{-5}^{+6}$ & $_{-0.019}^{+0.024}$ & $_{-0.006}^{+0.006}$ & $_{-0.004}^{+0.005}$ & $\pm0.004$ & $\pm 0.002$& $\pm 0.022$         \\
0.22    & $0.108^{+0.009}_{-0.010}$ & $0.5_{-0.1}^{+0.2}$ & $48_{-7}^{+7}$ ($118_{-9}^{+9}$)& $_{-0.006}^{+0.006}$ & $_{-0.002}^{+0.001}$ & $_{-0.004}^{+0.002}$ & $\pm 0.001$ & $_{-0.001}^{+0.002}$ & $\pm 0.005$         \\
0.46    & $0.136^{+0.009}_{-0.008}$ & $1.8_{-0.5}^{+0.6}$ & $230_{-13}^{+14}$ & $_{-0.006}^{+0.007}$ & $_{-0.002}^{+0.003}$ & $\pm 0.004$ & $\pm 0.001$ & $\pm 0.001$& $\pm 0.001$         \\
0.89    & $0.127^{+0.031}_{-0.015}$ & $3.7_{-0.9}^{+0.8}$ & $288_{-12}^{+11}$& $_{-0.006}^{+0.006}$ & $_{-0.006}^{+0.028}$ & $\pm 0.008$ & $\pm 0.006$ & $_{-0.002}^{+0.001}$& $\pm 0.007$         \\
2.33    & $0.173^{+0.044}_{-0.019}$ & $7.7_{-2.2}^{+3.2}$ & $377_{-14}^{+16}$& $_{-0.006}^{+0.006}$ & $_{-0.012}^{+0.042}$ & $\pm 0.008$ & $_{-0.007}^{+0.002}$ & $_{-0.001}^{+0.003}$& $\pm 0.010$         \\
3.91    & $0.236^{+0.055}_{-0.009}$ & $8.4_{-1.9}^{+2.4}$ & $318_{-15}^{+12}$ & $_{-0.004}^{+0.005}$ & $_{-0.007}^{+0.054}$ & $_{-0.002}^{+0.004}$ & $_{-0.003}^{+0.006}$ & $\pm 0.002$& $\pm 0.001$         \\
\hline\hline
\end{tabular}
\caption{Measured silicon ionization yield, Fano factor, and signal normalization, with uncertainties. The 220 eV normalization given without (with) parentheses is for the far (near) position of the ring detectors. The remaining columns provide the statistical (stat.) uncertainty and the systematic uncertainties from the uncertainty on the recoil energy, the neutron beam energy, the charge trapping and impact ionization probabilities (CT/II), the time of flight (TOF) cut, and (potential) deficiencies in modeling the Fano factor.}
\label{table:results}
\end{table*}

The systematic uncertainties are provided below.
\begin{enumerate}[topsep=0pt,itemsep=-1ex,partopsep=1ex,parsep=1ex]
    \item Recoil energy uncertainty: This has two contributions. A $\pm$1.3\% uncertainty arises from the total phonon energy scale calibration of the 0\,V mode data. For the LW detectors additional uncertainties come from the scale factors in Table~\ref{table:recoil_energy}.
    \item Neutron beam energy uncertainty: The central energy of the neutron beam as measured had a spread of $\pm 3$\keV. We vary the beam energy in the simulation and use the resulting spectra for the TOF cut and the fit model.
    \item Charge trapping and impact ionization uncertainties: These probabilities are varied by the uncertainty from Ref.~\cite{ren:2020}. They are varied conservatively such that when one probability is increased the other is decreased.
    \item Time of flight cut uncertainty: The neutron time of flight is correlated with their energy. The effect of the TOF selection is evaluated by choosing a wider (narrower) window selecting events within $\pm50$\% ($\pm20$\%) of the simulated TOF distribution maximum.
    \item Uncertainty in modeling the Fano factor: due to poor knowledge of the Fano factor, we perform a fit with the Fano factor fixed to one.
\end{enumerate}  
Each systematic uncertainty is evaluated at $1\sigma$ significance by fluctuating the corresponding parameter and performing the MCMC fit to 100 pseudoexperiments. These pseudoexperiments are generated by applying our detector response model to the simulated recoil energies with the nominal fit results. Each bin in the resulting distribution is then fluctuated by a Poisson random number, resulting in a single pseudoexperiment. To be conservative, we choose the deviation of the central yield value in each scenario from the original best fit yield plus the standard deviation of these fits as our estimate for the given systematic uncertainty. The systematic uncertainties are assigned two-sided asymmetric if the positive and negative fluctuated samples yield two-sided deviations, otherwise symmetric two-sided uncertainties are assigned with the largest deviation. All calculated systematic uncertainties are added in quadrature to obtain the total systematic uncertainty. For the LWs, the uncertainty on the recoil energy dominates because of the large position uncertainty. The statistical uncertainty dominates for the ring detectors.

\begin{figure}[htbp]
    \centering
    \includegraphics[width=0.46\textwidth]{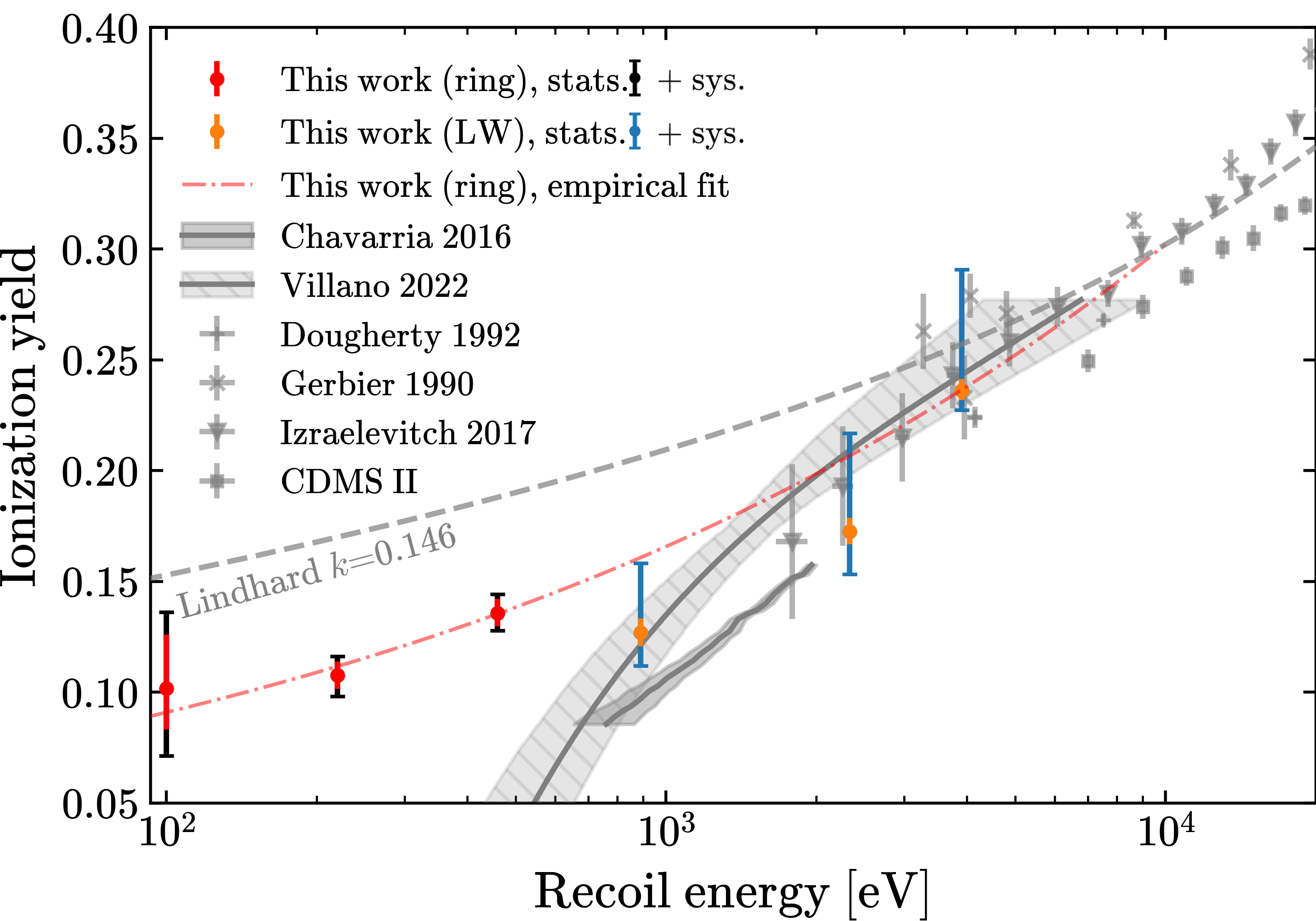}    
    \caption{The measured ionization yields, along with their statistical and total uncertainties and a fit with a power-law function. Also shown are data points from previous measurements \cite{agnese2018nuclear,gerbier1990measurement,dougherty1992measurements,chavarria2016measurement,izraelevitch2017measurement,Villano:2022}. The dashed line shows the Lindhard model with $k=$0.146~\cite{lewin1996yieldk}. 
    }
    \label{fig:results}
\end{figure}

The fit results and the uncertainties are provided in Table~\ref{table:results} and Fig.~\ref{fig:results}. The correlations among the fit parameters are found to be negligible. 
We provide a least-square fit to our results on the ring detectors with an empirically chosen power-law function $Y(E_r)=Y_{\mathrm{10keV}}\cdot (E_r/10000)^B$ that is constrained to go through the yield of Lindhard model at 10\,keV ($Y_{\mathrm{10keV}}$). The resulting best fit yields $B=0.261_{-0.011}^{+0.017}$, given $Y_{\mathrm{10keV}}=0.302$.

Our results show some tension with an earlier experiment using a photo-neutron source~\cite{chavarria2016measurement} at 890~eV. There is also tension at lower energies with the recent result using silicon neutron capture~\cite{Villano:2022}, which may be caused by the choice to fit to the Sorensen model~\cite{Sorensen:2015df} with a finite ionization threshold. Our results agree with the similar neutron-scattering setup~\cite{izraelevitch2017measurement} above 2\,keV.
Our measurement of the ionization yield of nuclear recoils in silicon is the first reaching down to 100\eV. The previously noted deviation from the Lindhard model extends down to 100\eV with no indication of an ionization production threshold. This latter fact is of great importance to rare event search experiments in semiconductor detectors.

\section{Acknowledgements}

\begin{acknowledgements}

The SuperCDMS collaboration gratefully acknowledges the Triangle Universities Nuclear Laboratory (TUNL) facility and its staff. Funding and support were received from the National Science Foundation, the U.S. Department of Energy (DOE), Fermilab URA Visiting Scholar Grant No.~15-S-33, NSERC Canada, the Canada First Excellence Research  Fund, the Arthur B. McDonald Institute (Canada),  the Department of Atomic Energy Government of India (DAE), the Department of Science and Technology (DST, India) and the DFG (Germany) - Project No.~420484612 and under Germany’s Excellence Strategy - EXC 2121 ``Quantum Universe" – 390833306. Femilab is operated by Fermi Research Alliance, LLC,  SLAC is operated by Stanford University, and PNNL is operated by the Battelle Memorial Institute for the U.S. Department of Energy under contracts DE-AC02-37407CH11359, DE-AC02-76SF00515, and DE-AC05-76RL01830, respectively. The TUNL accelerator is operated and maintained for the U.S. Department of Energy with grant DE-FG02-97ER41033. This research was enabled in part by support provided by SciNet (www.scinethpc.ca) and the Digital Research Alliance of Canada (www.alliancecan.ca).
\end{acknowledgements}
\bibliographystyle{apsrev4-1}
\bibliography{refs}

\end{document}